\documentclass{appolb}
\usepackage{graphicx}

\begin{document}
\title{Spin-thermal shear coupling in relativistic nuclear collisions %
\thanks{Presented at the XXIXth International Conference on Ultra-relativistic Nucleus-Nucleus Collisions (Quark Matter 2022)}%
}
\author{M. Buzzegoli
\address{Department of Physics and Astronomy, Iowa State University,\\ Ames, Iowa 50011, USA}
\\[3mm]
{F. Becattini
\address{Universit\`a di Firenze and INFN Sezione di Firenze,\\ Via G. Sansone 1, I-50019 Sesto Fiorentino (Florence), Italy}
}
\\[3mm]
{G. Inghirami
\address{GSI Helmholtzzentrum f\"ur Schwerionenforschung GmbH,\\Planckstr. 1, 64291 Darmstadt , Germany}
}
\\[3mm]
{I. Karpenko
\address{Faculty of Nuclear Sciences and Physical Engineering, Czech Technical University in Prague, B\v rehov\'a 7, 11519 Prague 1, Czech Republic}
}
\\[3mm]
A. Palermo
\address{Universit\`a di Firenze and INFN Sezione di Firenze,\\ Via G. Sansone 1, I-50019 Sesto Fiorentino (Florence), Italy\\
and Institut f\"ur Theoretische Physik, Johann Wolfgang Goethe-Universit\"at, Max-von-Laue-Straße 1, D-60438 Frankfurt am Main, Germany}
}
\maketitle
\begin{abstract}
The spin polarization measurements of particles emitted in heavy-ion collisions have opened the possibility for new phenomenological investigations of spin physics in relativistic fluids. The theoretical predictions of global polarization are in agreement with the data, but consistent discrepancies stand out for the local polarization. We show that the covariant theory of relativistic quantum fluids at local equilibrium implies an additional, non-dissipative contribution to the spin polarization vector, which is proportional to the thermal shear, which has been previously overlooked. This additional contribution, together with an improved approximation in the expansion of the local equilibrium density operator, restores the quantitative agreement between the theoretical predictions and the experimental data.
\end{abstract}
  
\section{Introduction}
Spin polarization in the Quark Gluon Plasma created in relativistic heavy ion collisions was first reported in \cite{star,Adam:2018ivw}.
The measured global (i.e., integrated over all momenta) spin polarization of $\Lambda$ hyperons is in quantitative
agreement \cite{Becattini:2020ngo} with the predictions
based on local thermodynamic equilibrium, which implies a relation between spin and thermal vorticity~\cite{Becattini:2013fla}.
However, those predictions are not able to reproduce the local (i.e., as a function of momentum) polarization measured
in \cite{Adam:2019srw,Niida:2018hfw} for Au+Au collisions at 200 GeV. In particular, the sign of the longitudinal
component of spin polarization $(P^z)$ as a function of the azimuthal angle is opposite to the model predictions \cite{Becattini:2020ngo}.

In this work, which is largely based on ref. \cite{Becattini:2021suc,Becattini:2021iol},
we will show that the inclusion of the effect of
the thermal shear and an improved approximation of the statistical operator restore the quantitative agreement between hydrodynamic model
predictions and local polarization measurements.

\section{Spin polarization}
We start by deriving the spin polarization vector for a relativistic fluid close to the local thermal equilibrium.
The mean spin vector of a $\Lambda$ hyperon is obtained from \cite{Becattini:2020sww}
\begin{equation}
\label{eq:SpinPolFormula}
S^\mu(p)= \frac{1}{2}\frac{\int_{\Sigma_{FO}} {\rm d}\Sigma\cdot p\; {\rm tr}_4\left[\gamma^\mu \gamma^5 W_+(x,p) \right] }
 {\int_{\Sigma_{FO}} {\rm d}\Sigma\cdot p\; {\rm tr}_4 \left[W_+(x,p)\right]},
\end{equation}
where $\Sigma_{FO}$ is the Freeze-Out hadronization 3D hypersurface, and $ W_+(x,p)$ denotes the future time-like part of the Wigner function.
For a weakly interacting hadron, we can use the Wigner function of the free Dirac field:
\begin{equation}
\label{eq:WignerFunc}
W(x,k)_{AB} = \frac{1}{(2\pi)^4} \! \int \!{\rm d}^4 y\, {\rm e}^{-{\rm i} k \cdot y}
	\langle : \bar{\Psi}_B (x +y/2) \Psi_A (x-y/2) : \rangle,
\end{equation}
where the symbol $\langle \widehat{X} \rangle = {\rm tr} \left( \widehat{\rho}\, \widehat{X} \right)$ denotes a thermal average
with the statistical operator $\widehat{\rho}$.

In the hydrodynamic picture of the QCD plasma, the statistical operator is assumed to be the local equilibrium density operator
specified by the initial conditions \cite{Becattini:2019dxo}. Neglecting the dissipative effects, it is given with a good
approximation (corresponding to ideal dissipationless hydrodynamic) by its local equilibrium form:
\begin{equation}
\label{eq:LEStatOper}
\widehat{\rho}\simeq \widehat{\rho}_{{\rm LE}}
	= \frac{1}{Z} \exp\left[ -\int_{\Sigma_{FO}}\!\! {\rm d}\Sigma_\mu
	\widehat{T}^{\mu\nu}\beta_\nu
	\right],
\end{equation}
where $\beta = ( 1 / T ) u$ is the four-temperature vector
and $\widehat{T}^{\mu\nu}$ is the stress-energy tensor.

The Wigner function (\ref{eq:WignerFunc}) resulting from the statistical operator (\ref{eq:LEStatOper})
is obtained by taking advantage of the separation of scales in the hydrodynamic regime.
In fact, since $\beta$ is slowly varying compared to the microscopic length scales, one can Taylor expand
it around the point $x$:
\begin{equation}
\label{eq:TaylorBeta}
\beta_\nu(y) \simeq \beta_\nu(x) + \partial_\lambda \beta_\nu(x) (y-x)^\lambda+\cdots\, .
\end{equation}
Stopping at the first order in derivatives, one can approximate the statistical operator as
\begin{equation}
\label{eq:ApproxHydro}
\widehat{\rho}_{LE} \simeq \frac{1}{Z} \exp \left[ - \beta_\nu(x) \widehat{P}^\nu
	+ \frac{1}{2} \varpi_{\mu\nu}(x) \widehat{J}^{\mu\nu}_x  -\frac{1}{2} \xi_{\mu\nu}(x) \widehat{Q}^{\mu\nu}_x +\cdots\right]
\end{equation}
where $\widehat{P}^\nu$ is the total four-momentum, $\widehat{J}^{\mu\nu}_x$ is the total
angular momentum operator, and
$
\widehat{Q}^{\mu\nu}_x =\! \int \!{\rm d} \Sigma_\lambda \left[ (y-x)^\mu \widehat{T}^{\lambda\nu}(y) + 
  (y-x)^\nu \widehat{T}^{\lambda\mu}(y)\right],
$
while
\begin{equation}
\varpi_{\mu\nu}=-\frac{1}{2}\left(\partial_\mu\beta_\nu - \partial_\nu\beta_\mu \right),\quad
\xi_{\mu\nu}=\frac{1}{2}\left(\partial_\mu\beta_\nu + \partial_\nu\beta_\mu \right)
\end{equation}
are respectively the thermal vorticity and the thermal shear.
Using the linear response theory, the mean spin vector (\ref{eq:SpinPolFormula}) result in \cite{Becattini:2021suc}
\begin{equation}
\label{eq:SpinPolFirstOrder}
S^\mu(p) = - \epsilon^{\mu\rho\sigma\tau} p_\tau 
  \frac{\int_{\Sigma_{FO}} {\rm d} \Sigma \cdot p \; n_F (1 -n_F) 
  \left[ \varpi_{\rho\sigma} + 2\, \hat t_\rho \frac{p^\lambda}{\varepsilon} \xi_{\lambda\sigma} \right]}
  { 8m \int_{\Sigma_{FO}} {\rm d} \Sigma \cdot p \; n_F},
\end{equation}
with $n_F=\left({\rm e}^{\beta\cdot p-\zeta}+1 \right)^{-1}$ and $\hat{t}$ is the time direction in the laboratory frame.
The first term is the spin polarization induced by thermal vorticity~\cite{Becattini:2013fla}, and
the second is the spin polarization induced by thermal shear \cite{Becattini:2021suc,Liu:2021uhn}.
The thermal shear contribution was neglected in previous analysis, but its presence was also confirmed
in later studies \cite{Liu:2021nyg,Yi:2021ryh}.

The impact of shear induced polarization in heavy-ion collisions is studied in
ref. \cite{Becattini:2021iol,Yi:2021ryh,Fu:2021pok,Ryu:2021lnx,Florkowski:2021xvy,Sun:2021nsg,Alzhrani:2022dpi,Wu:2022mkr}.
It was found that the thermal shear contribution in (\ref{eq:SpinPolFirstOrder}) helps in reducing
the discrepancies with the data taken for Au+Au collisions at 200 GeV, but does not recover a full
quantitative agreement and that local polarization can be very sensitive to the equation of states,
viscosities, and the freeze-out temperature.

\section{Isothermal equilibrium}
\begin{figure}[htb]
\centerline{%
\includegraphics[width=5.5cm]{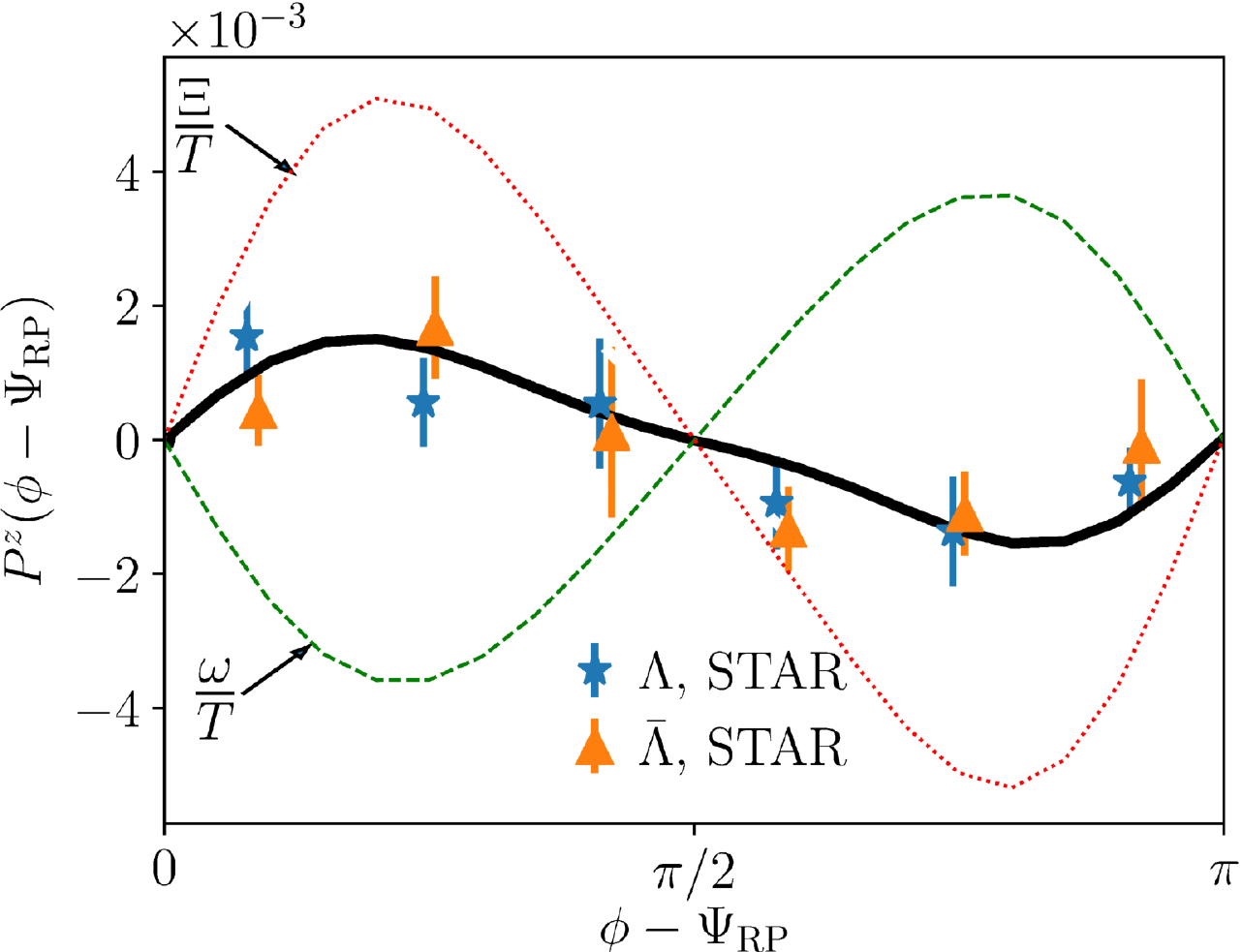}}
\caption{Longitudinal components of $\Lambda$ polarization ($P^z$) at ILE as
a function of the azimuthal angle for a decoupling temperature of $T_{\rm FO}=150$ MeV~\cite{Becattini:2021iol}.
The calculations are done with averaged MC Glauber IS corresponding to 20-60\% central Au-Au collisions at
200 GeV RHIC energy. Experimental data points are taken from \cite{Adam:2019srw}.}
\label{Fig:PzAngle}
\end{figure}

For relativistic nuclear collision at very high
energy, the chemical potentials are negligible,
and temperature is the only effective
intensive variable. Then, the decoupling occurs at a
constant
temperature, and $\Sigma_{FO}$ is an isothermal hypersurface. It follows that, at very high energy,
the statistical operator~(\ref{eq:LEStatOper}) assumes the form of Isothermal Local Equilibrium (ILE)
obtained by moving the constant temperature out of the integral sign:
\begin{equation}
\label{eq:ILEStatOper}
\widehat{\rho}_{\rm ILE} = \frac{1}{Z} \exp\left[
-\int_{\Sigma_{FO}}\!\! {\rm d}\Sigma_\mu \widehat{T}^{\mu\nu}\left( \frac{u_\nu}{T_{\rm FO}}\right) \right]
=\frac{1}{Z} \exp\left[-\frac{1}{T_{\rm FO}}\int_{\Sigma_{FO}}\!\! {\rm d}\Sigma_\mu \widehat{T}^{\mu\nu}u_\nu \right].
\end{equation}
We now evaluate the mean spin polarization (\ref{eq:SpinPolFormula}) using (\ref{eq:ILEStatOper}).
It is clear that, in this situation, we do not need to Taylor expand the whole $\beta$ field as done
in Eq. (\ref{eq:TaylorBeta}), but only the fluid velocity:
$u_\nu(y) \simeq u_\nu(x) + \partial_\lambda u_\nu(x) (y-x)^\lambda+\cdots$.
The result is the following improved formula for spin polarization at high energy \cite{Becattini:2021iol}:
\begin{equation}
\label{eq:ILEPol}
S_{\rm ILE}^\mu(p) = - \epsilon^{\mu\rho\sigma\tau} p_\tau 
  \frac{\int_{\Sigma_{FO}} {\rm d} \Sigma \cdot p \; n_F (1 -n_F) 
  \left[ \omega_{\rho\sigma} + 2\, \hat t_\rho \frac{p^\lambda}{\varepsilon} \Xi_{\lambda\sigma} \right]}
  { 8m T_{\rm FO} \int_{\Sigma_{FO}} {\rm d} \Sigma \cdot p \; n_F},
\end{equation}
with $\omega$ and $\Xi$, the kinematic vorticity and shear:
\begin{eqnarray}
\label{eq:Decompositions}
\omega_{\rho\sigma} =& \frac{1}{2} \left(\partial_\sigma u_\rho - \partial_\rho u_\sigma \right)
	= A_\rho u_\sigma -  A_\sigma u_\rho + \frac{1}{2}\epsilon_{\rho\sigma\mu\nu}\omega^\mu u^\nu,\\
\label{eq:Decompositions2}
\Xi_{\rho\sigma} =&  \frac{1}{2} \left(\partial_\sigma u_\rho + \partial_\rho u_\sigma \right)
	= \frac{1}{2} \left( A_\rho u_\sigma +
   A_\sigma u_\rho \right) + \sigma_{\rho\sigma} + \frac{1}{3} \theta \Delta_{\rho\sigma},
\end{eqnarray}
where $\Delta_{\mu\nu}=g_{\mu\nu}-u_\mu u_\nu$, $A=u\cdot \partial u$ is the acceleration field, $\sigma$ is the shear tensor
$\sigma_{\mu\nu} = \frac{1}{2} (\nabla_\mu u_\nu + \nabla_\nu u_\mu) - \frac{1}{3} \Delta_{\mu\nu} \theta$,
and $\theta = \nabla \cdot u$, with $\nabla_\mu=\partial_\mu-u_\mu \partial\cdot u$. By comparing Eq.~(\ref{eq:ILEPol})
with Eq.~(\ref{eq:SpinPolFirstOrder}) we see that if we do not use the ILE at high energies we are evaluating the spin
polarization with an error proportional to the temperature gradients. Notice however that temperature gradients are not
being neglected in the ILE as they are still non-vanishing even though orthogonal to the hypersurface.

\begin{figure}[htb]
\centerline{%
\includegraphics[width=9cm]{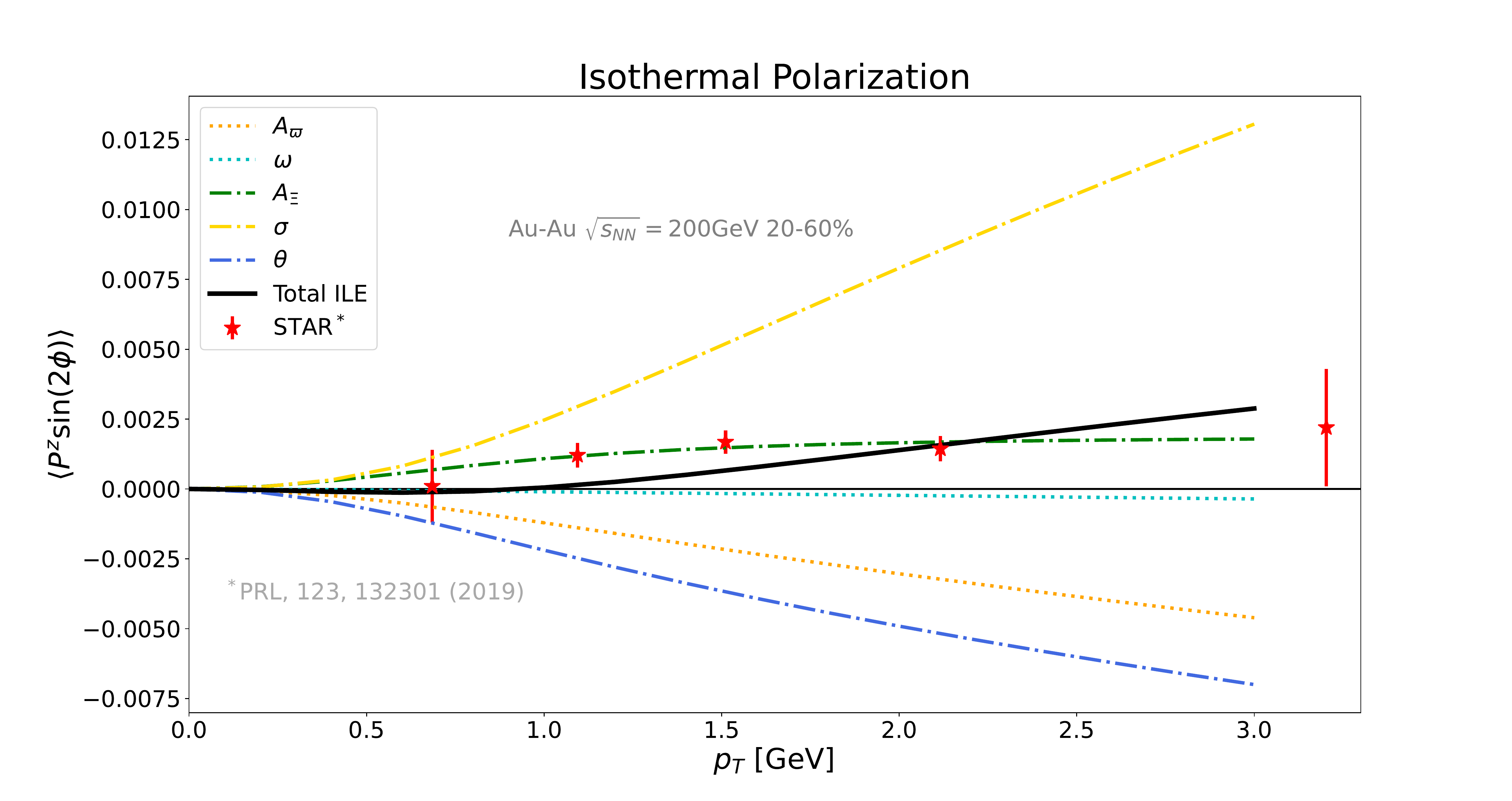}}
\caption{Contributions to the quadrupole longitudinal components of $\Lambda$ polarization at ILE
stemming from kinematic vorticity $\omega$, shear tensor $\sigma$, acceleration (from vorticity $A_\varpi$ and shear $A_\Xi$) and
expansion rate $\theta$, see Eqs.~(\ref{eq:Decompositions}) and~(\ref{eq:Decompositions2}) .}
\label{Fig:PzComponents}
\end{figure}
In Fig.~\ref{Fig:PzAngle} we report the model predictions of $P^z$ obtained with the Eq.~(\ref{eq:ILEPol})
at $T_{\rm FO}=150$ MeV. The inclusion of thermal shear and the use of ILE is then capable
of reproducing the experimental data.
In Fig.~\ref{Fig:PzComponents}, we also report the second-order
Fourier harmonic coefficient of longitudinal polarization as a function of transverse momentum. Future investigations
are needed to study the decay contribution to spin polarization, the dependence on viscosities and on initial
conditions, and to make predictions at other energies.

\textbf{Acknowledgments.} M.B. is supported by the US Department of Energy under Grant No. DE-FG02-87ER40371.


\end{document}